\documentclass[aps,prb,superscriptaddress,reprint,amssymb,longbibliography]{revtex4-2}
\usepackage[english]{babel}
\usepackage{graphicx}
\usepackage[utf8]{inputenc}
\usepackage{verbatim}
\usepackage[version=3]{mhchem}
\usepackage{amsmath}  
\usepackage{xcolor}
\usepackage{upgreek}
\usepackage{bm}
\usepackage{amssymb}
\usepackage{braket}
\usepackage{microtype} 
\usepackage{textcomp}
\usepackage{hyperref}
\usepackage{cleveref}

\begin{document}

\title{Unified Semiclassical Theory of Nonlinear Hall Effect\texorpdfstring{:\\}{: }Bridging Ballistic and Diffusive Transport Regimes}

\author{X. Y. Liu}
\affiliation{National Laboratory of Solid State Microstructures, School of Physics,
and Collaborative Innovation Center of Advanced Microstructures, Nanjing University, Nanjing 210093, China}

\author{H. Z. Liao}
\affiliation{National Laboratory of Solid State Microstructures, School of Physics,
and Collaborative Innovation Center of Advanced Microstructures, Nanjing University, Nanjing 210093, China}

\author{G. Y. Qi}
\affiliation{College of Electrical Engineering, Zhejiang University of Water Resources and Electric Power, Hangzhou 310018, China}

\author{H. Geng}
\email{genghao@nuaa.edu.cn}
\affiliation{College of Physics, Nanjing University of Aeronautics and Astronautics, Nanjing 211106, China}
\affiliation{Key Laboratory of Aerospace Information Materials and Physics (NUAA), MIIT, Nanjing 211106, China}

\author{L. Sheng}
\email{shengli@nju.edu.cn}
\affiliation{National Laboratory of Solid State Microstructures, School of Physics,
and Collaborative Innovation Center of Advanced Microstructures, Nanjing University, Nanjing 210093, China}

\author{D. Y. Xing}
\affiliation{National Laboratory of Solid State Microstructures, School of Physics,
and Collaborative Innovation Center of Advanced Microstructures, Nanjing University, Nanjing 210093, China}


\date{\today}

\begin{abstract}
The nonlinear Hall effect has attracted considerable attention and undergone extensive investigation in recent years. However, theoretical studies addressing size-dependent effects remain largely unexplored. In this work, we establish a unified semiclassical framework based on the Boltzmann transport equation, incorporating generalized boundary conditions to bridge the ballistic and diffusive transport regimes.
Our analysis reveals that the nonlinear Hall effect arises from the combined action of two distinct mechanisms: the Berry curvature dipole and the Fermi-surface integral of Berry curvature. 
Furthermore, we investigate the Hall effect in topological crystalline insulators (TCIs), elucidating that the size dependence originates from competition between the two transport mechanisms.
By connecting the two distinct regimes, our theoretical framework provides a comprehensive understanding of the nonlinear Hall effect in finite-sized systems, offering both fundamental insights and a useful analytical tool for more size-dependent investigations.
\end{abstract}

\maketitle
\section{Introduction}

The Hall effect was first discovered by Edwin Hall in 1879 \cite{Hall1879}, describing the emergence of a transverse voltage in conductors under mutually perpendicular electric and magnetic fields. Subsequent studies revealed that in certain materials with spontaneous magnetization, a transverse voltage could emerge even without an external magnetic field. This phenomenon, known as the anomalous Hall effect \cite{karplusHallEffectFerromagnetics1954,nagaosaAnomalousHallEffect2010}, bridges quantum geometric Berry phase effects with macroscopic transport phenomena. With advancements in experimental techniques, the quantum Hall effect was discovered in two-dimensional(2D) electron systems under low temperature and strong magnetic field conditions, where Hall conductivity exhibits quantized plateaus \cite{Klitzing1980,laughlinQuantizedHallConductivity1981,tsuiTwoDimensionalMagnetotransportExtreme1982a}. Recent studies show that magnetized two-dimensional topological insulators can exhibit quantized Hall conductivity without an external magnetic field—a phenomenon dubbed the quantum anomalous Hall effect \cite{liuQuantumAnomalousHall2008, changExperimentalObservationQuantum2013, yuQuantizedAnomalousHall2010, nomuraSurfaceQuantizedAnomalousHall2011, gaoQuantumAnomalousHall2020, yoshimiQuantumHallStates2015}. However, these conventional Hall effects require broken time-reversal symmetry. In contrast, the nonlinear Hall effect emerges in systems with broken inversion symmetry but preserved time-reversal symmetry, originating from quantum geometric properties--specifically the Berry curvature dipole--that arise from the crystalline inversion symmetry breaking. In such systems, the positive and negative Berry curvatures in different momentum regions are separated to generate Berry curvature dipole. The nonlinear Hall effect was first observed in bilayer $\text{WTe}_2$~\cite{maObservationNonlinearHall2019a}, subsequently were confirmed in multilayer $\text{WTe}_2$~\cite{kangNonlinearAnomalousHall2019a, xiaoBerryCurvatureMemory2020}, nonmagnetic Weyl-Kondo semimetal $\text{Ce}_3\text{Bi}_4\text{Pd}_3$~\cite{dzsaberGiantSpontaneousHall2021a}, ferroelectric Weyl semimetal $\text{Pb}_{1-x}\text{Sn}_x\text{Te}$~\cite{zhangGiantBerryCurvature2022}, Weyl semimetal $\text{TaIrTe}_4$~\cite{kumarRoomtemperatureNonlinearHall2021}, $\text{NbP}$\cite{zhangTerahertzDetectionBased2021}, and topological insulators including $\text{MnBi}_2\text{Te}_4$~\cite{dengQuantumAnomalousHall2020}, $\text{Pb}_{1-x}\text{Sn}_x\text{Te}$~\cite{nishijimaFerroicBerryCurvature2023} and $\text{ZrTe}_5$~\cite{wangNoncentrosymmetricTopologicalPhase2024}, as well as heterointerfaces such as $\text{WSe}_2/\text{SiP}$ interface~\cite{duanBerryCurvatureDipole2023}, $\text{LaAlO}_3/\text{SrTiO}_3$ interface~\cite{lesneDesigningSpinOrbital2023}. Remarkably, the nonlinear Hall effect can even exist at room temperature in $\text{TaIrTe}_4$~\cite{kumarRoomtemperatureNonlinearHall2021}, $\text{NbP}$~\cite{zhangTerahertzDetectionBased2021}, $\text{BaMnSb}_2$~\cite{minStrongRoomtemperatureBulk2023}, and elemental semiconductor tellurium~\cite{chengGiantNonlinearHall2024}.

In theoretical studies, the nonlinear Hall effect was initially formulated within the semiclassical framework of Berry curvature dipole. Subsequently, the Magnus Hall effect~\cite{papajMagnusHallEffect2019} was proposed for the ballistic transport regime. This phenomenon arises from the quantum Magnus effect: a rotating Bloch electron wave packet, when subjected to an electric field, experiences an anomalous transverse velocity.
Significantly, in the ballistic limit, the Magnus Hall conductance directly reflects the Berry curvature distribution on the Fermi surface. Both the Magnus Hall effect (in the ballistic regime) and the nonlinear Hall effect (in the diffusive regime) demonstrate intimate connections to Berry curvature, revealing a fundamental relationship between these apparently distinct transport phenomena.

In this work, we establish the intrinsic relationship between these two effects and develop a unified semiclassical conductance formula for the nonlinear Hall effect that bridges diffusive and ballistic transport regimes. By introducing a position-dependent electric field with generalized boundary conditions, we derive a universal conductance relation through the Boltzmann transport equation (BTE) capable of describing electron transport in finite-sized systems across the entire diffusive-to-ballistic regimes. The derived formula naturally reduces to the nonlinear Hall effect in the diffusive limit and simplifies to the Magnus Hall effect in the ballistic limit. Furthermore, we apply this approach to study topological crystalline insulator (TCIs) systems as an explicit example. Due to the interplay of different mechanisms, the nonlinear Hall conductance and the Fermi energy corresponding to its peak display a shift as the system size or mean free path varies. 

The rest of this article is organized as follows. In Sec.~\ref{sec.method}, we present the derivation of the universal nonlinear conductance formula through the semiclassical Boltzmann transport framework, along with the formulas for the diffusive and ballistic limits. In Sec.~\ref{sec.result}, we derive the approximate analytical solution for the conductivity of TCIs in the diffusive limit and numerically calculated the conductance at different sizes of system. Finally, in Sec.~\ref{sec.conclusion}, we have summarized and  discussed the results.

\begin{figure}[htbp]
    \centering
    \includegraphics[width=\linewidth]{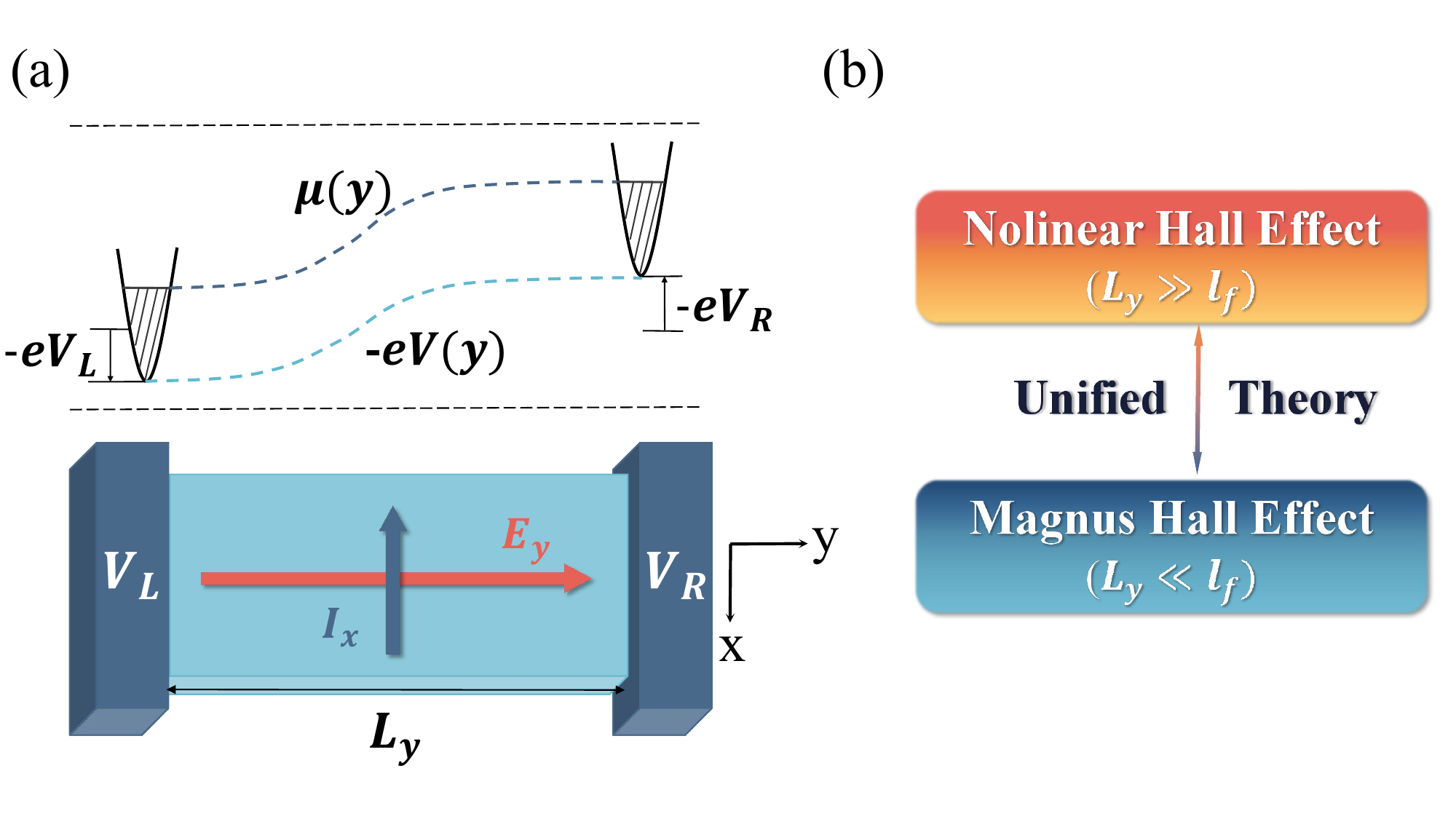}
    \caption{(a) Schematics of finite-sized system  with large electronic reservoirs connected on both sides. An electric field is applied in the y direction between the source and drain regions. The bottom of the energy band depends on the bias voltage, where $ \mu(y)$ and $-eV(y)$ correspond to the chemical potential and electrical potential, respectively. (b) Unified theory bridging diffusive-to-ballistic transport regimes, where $L_y$ is the length of the sample and $l_f$ is the electron mean free path.
}
    \label{Fig1}
\end{figure}

\section{Unified Semi-classical Formula for Nonlinear Hall Effect}\label{sec.method}
In this section, we establish a unified semiclassical theory for the nonlinear Hall effect. We consider a two-dimensional conductor with finite length $L_y$ in the $y$-direction and infinite extent in the $x$-direction. The sample is connected to two large electronic reservoirs located in regions $y<0$ and $y>L_y$, as illustrated in Fig.~\ref{Fig1}(a). 
When a voltage bias is applied between two reservoirs, it generates an electric field $\mathbf{E}$ which confined within the sample and drives electron transport opposite to the direction of the electric field, generating a drift current.
Moreover, due to the different voltage biases of the reservoirs, a spatially inhomogeneous electron density distribution emerges. 
This inhomogeneous distribution gradient generates an additional contribution to the electron transport, known as the diffusion current. To describe both drift and diffusion contributions to the electron current, we employ the BTE within the relaxation time approximation
\begin{equation}\label{eq.btefull}
\frac{\partial f}{\partial t} + \dot{\mathbf{r}} \cdot \nabla_{\mathbf{r}} f + \dot{\mathbf{k}} \cdot \nabla_{\mathbf{k}} f = -\frac{f - \bar{f}}{\tau_0}~,
\end{equation}
where $f\left(\mathbf{k},\mathbf{r}\right)$ denotes the nonequilibrium distribution function. In the steady state, the first term on the left-hand side of Eq.~\eqref{eq.btefull} vanishes, namely $\frac{\partial f}{\partial t}=0$.
The spatial gradient term $\dot{\mathbf{r}} \cdot \nabla_{\mathbf{r}} f$ describes diffusion "force" due to the distribution function inhomogeneity, giving rise to the diffusion current. The momentum gradient term $\dot{\mathbf{k}} \cdot \nabla_{\mathbf{k}} f$ represents the electric-field-driven drift of carriers, generating the drift current. The electron dynamics are governed by the semiclassical equations of motion~\cite{Sundaram1999,Xiao2010}:
\begin{equation}\label{eq.rk}
 \dot{\mathbf{r}}=\frac{1}{\hbar}\nabla_{\mathbf{k}}\epsilon_\mathbf{k}-\dot{\mathbf{k}}\times\mathbf{\Omega}~,\ \ \dot{\mathbf{k}}=-\frac{e}{\hbar}\mathrm{\mathbf{E}}~,
\end{equation}
where $\mathbf{\Omega}$ denotes the Berry curvature, $\mathbf{r}$ and $\mathbf{k}$ are the center position and momentum of the electron wave packet, respectively. $-e$ is the electron charge, $\mathbf{E}$ is the external electric field.

The right-hand side of Eq.~\eqref{eq.btefull} represents the collision process, which describes the relaxation of the nonequilibrium distribution function towards the local equilibrium state.
Here, we assume that the quantum phase coherence length $l_\phi$ (primarily determined by inelastic electron-phonon scattering processes at finite temperatures) is much shorter than the electron mean free path $l_f$ (characterized by electron-impurity scattering with relaxation time $\tau_0$). Under this condition, the quantum interference effects can be neglected.
While decoherence is mainly governed by inelastic events, the relaxation towards a locally isotropic state in momentum space is driven by all scattering processes that change the electron's momentum. This includes both elastic scattering off static impurities (which preserves phase information but randomizes momentum) and the aforementioned inelastic scattering. Consequently, the nonequilibrium distribution function loses its directional dependence, resulting in a local equilibrium distribution function $\bar{f}$ that is dependent on the energy $\epsilon_\mathbf{k}$, the electric potential energy $-eV(\mathbf{r})$, the chemical potential $\mu(\mathbf{r})$, and the temperature $T$. And the local equilibrium distribution function $\bar{f}$ takes the equilibrium Fermi-Dirac distribution form 
\begin{equation}\label{eq.fbar}
\bar{f}\left(\epsilon_k,\mathbf{r}\right) =\frac{1}{e^{\tfrac{\epsilon_\mathbf{k}-e V(\mathbf{r})-\mu(\mathbf{r})}{k_B T}}+1}~.
\end{equation}
where $-eV(\mathbf{r})$ and $\mu(\mathbf{r})$ have been corrected for electron energy and Fermi energy, respectively.~\cite{Datta2000NanoscaleDeviceModeling,Datta2005QuantumTransportAtom,Datta2018LessonsNanoelectronicsNew}. Due to the limited transverse size of the system, the chemical potential exhibits spatial dependence, and its numerical distribution is influenced by the local potential energy.

In our specific setup as shown in Fig.~\ref{Fig1}(a), a voltage bias is applied in the $y$-direction, resulting in an electric field that is approximately confined within the sample and oriented along the $y$-direction. Thus,
\begin{equation}\label{eq.Efield}
    \mathbf{E}=E_y\mathbf{\hat{y}}=-\frac{\partial V(y)}{\partial y}\mathbf{\hat{y}}~.
\end{equation}
Here, $V_L$ and $V_R$ denote the voltages at the left and right ends of the sample, and $V(y)$ is the electric potential in the sample.
Neglecting the voltage drop at the contacts' interfaces between the sample and the reservoirs, which is reasonable when the decoherence length $l_\phi$ is much shorter than the sample size $L_y$~\cite{Datta2018LessonsNanoelectronicsNew,Datta2005QuantumTransportAtom}, we obtain the following relations:
\begin{equation}\label{eq.EV}
    V_L-V_R=\int_{0}^{L_y}E_ydy~, \quad V(y)=\int_{0}^{y}E_ydy~.
\end{equation}
Moreover, the non-equilibrium distribution function is approximately spatially solely dependent on the coordinate $y$ and is uniform along the $x$-direction. Overall, the original BTE from Eq.~\eqref{eq.btefull} can be simplified as
\begin{equation}\label{eq.btesim}
v_y \frac{\partial f}{\partial y}- e v_y \frac{\partial V\left(y\right)}{\partial y} \left( -\frac{\partial f}{\partial \epsilon_k}\right) = -\frac{f - \bar{f}}{\tau_0}~,
\end{equation}
where $v_y=\frac{1}{\hbar}\frac{\partial\epsilon_k}{\partial k_y}$ is the group velocity of electrons and $\mu(\mathbf{r})=\mu(y)$.

In the weak-field condition which is characterized by small bias voltage and gradually slowly varying electric potential, we expand the nonequilibrium distribution function to first order around the equilibrium distribution function~\cite{Sheng1994PathintegralApproachQuasiclassical,Sheng1997QuasiclassicalApproachMagnetotransport,Sheng1998BoltzmannEquationSpindependent}, yielding
\begin{equation}\label{eq.fexpand}
f\left(\mathbf{k},y\right) =f_0 + \left(-\frac{\partial f_0}{\partial \epsilon_k}\right)[e V(y) + {g}(\mathbf{k},y)]~,
\end{equation}
where $f_0 = 1/(e^{(\epsilon_{k} - E_{F})/k_B T} + 1)$ is the Fermi-Dirac equilibrium distribution and $g(\mathbf{k},y)$ is a non-equilibrium distribution function related to the chemical potential and the electric potential energy~\cite{Boettcher2021}. Carrier transport is considered to be driven by the so-called electrochemical potential, which is the sum of the chemical potential and the electric potential energy~\cite{Boettcher2021}. Next, we expand the local distribution function to the linear order as
\begin{equation}\label{eq.ldos}
    \bar{f}\left(\epsilon_k,y\right) =f_0 +\left(-\frac{\partial f_0}{\partial \epsilon_k}\right)[e V(y)+\mu(y)-E_F ]~.
\end{equation}
Taking Eq.~\eqref{eq.fexpand} and Eq.~\eqref{eq.ldos} into Eq.~\eqref{eq.btesim}, we can eliminate the second term on the left-hand side of Eq.~\eqref{eq.btesim} and obtain
\begin{equation}\label{eq.bteg}
    v_y \frac{\partial g}{\partial y} = -\frac{g - \bar{g}}{\tau_0}~,
\end{equation}
where we denote
\begin{equation}\label{eq.barg}
    \bar{g}(y)=\mu(y) - E_F~.  
\end{equation}

Though Eq.~\eqref{eq.bteg} is much simpler compared with Eq.~\eqref{eq.btefull}, it cannot be solved directly. Because the electric potential $V(y)$ (or $E_y$) and local chemical potential $\mu(y)$ are unknown, we generally need to solve the Poisson-Boltzmann equation self-consistently~\cite{Sharp1990CalculatingTotalElectrostatic,Lamm2003PoissonBoltzmannEquation}.
However, in our approach, we employ the particle number conservation ansatz over the entire momentum space $\mathbf{k}$ at each position $\mathbf{r}$, which allows us to bypass the need for solving the Poisson equation. This condition is expressed as
\begin{equation}
   \int f( \mathbf{k},y )d^{D}k=\int \bar{f}( \mathbf{k},y )d^{D}k~,
\end{equation}
and is equivalent to
\begin{equation}
   \int g( \mathbf{k},y ) \left(-\frac{\partial f_0}{\partial \epsilon_k}\right)d^{D}k=\int \bar{g}( y ) \left(-\frac{\partial f_0}{\partial \epsilon_k}\right)d^{D}k~,
\end{equation}
where $D=2$ for the two-dimensional case, and this method can be readily extended to one- and three-
dimensions as well~\cite{gengUnifiedSemiclassicalApproach2016}.
The electrical current density can be written as
$$
j_y = \int v_y f\left(\mathbf{k},y\right)  d^2k\equiv \int v_y g\left(\mathbf{k},y\right)\left(-\frac{\partial f_0}{\partial \epsilon_k}\right)  d^2k~.
$$
This particle number conservation ansatz directly leads to the current conservation directly. Combining Eq.~\eqref{eq.bteg}, we can obtain
\begin{equation}\label{eq.Iyconser}
    \frac{\partial j_y}{\partial y}=0~.
\end{equation}

In the two reservoirs, the distribution function is in equilibrium. When bias voltages are applied, the Fermi levels of the two reservoirs are equal to the bias voltages (see Fig.~\ref{Fig1}(a)) respectively, which is the result from the charge neutrality condition in the reservoirs~\cite{Buttiker1995ChargeCurrentConserving,Christen1996GaugeinvariantNonlinearElectrica,Zou2024NonreciprocalBallisticTransport}. The conditions are given by
\begin{equation}\label{eq.muV}
    \mu_L= E_F- eV_L \quad \text{and}\quad \mu_R= E_F- eV_R~.  
\end{equation}
At the boundaries of the sample, we assume that electron modes are injected perfectly from the reservoirs without any reflection.
Therefore, we have
\begin{align}\label{eq.gboundary}
 \begin{cases}
        g\left(\mathbf{k},y=0^+\right)=-{\mathrm{eV}}_L\equiv g_L, &~v_y(\mathbf{k})>0 \\
        g\left(\mathbf{k},y=L_y-0^+\right)=-{\mathrm{eV}}_R\equiv g_R, &~v_y(\mathbf{k})<0~.
    \end{cases}
\end{align}
Note that $\bar{g}(y)$ is a function of the coordinate $y$, since the linear approximation holds well under two limits, we assume~\cite{gengUnifiedSemiclassicalApproach2016}
\begin{align}\label{eq.gbarlinear}
    \bar{g}\left(y\right)&=m+ny~,\\
    m&=\frac{g_LL_y+\left(g_L+g_R\right)l_f}{L_y+2l_f}~,\\
    n&=-\frac{g_L-g_R}{L_y+2l_f}~.
\end{align}
Combining Eq.~\eqref{eq.bteg} with boundary conditions Eq.~\eqref{eq.gboundary}, we can obtain the solution of $g\left(\mathbf{k},y\right)$ as
\begin{widetext}
  \begin{align}
g(\mathbf{k},y) =
\begin{cases}
    g_L e^{-\frac{y}{v_y\tau_0}} + \int_{0}^{y} e^{-\frac{y-\xi}{v_y\tau_0}} \frac{1}{v_y\tau_0} \bar{g}(\xi) \,\mathrm{d} \xi, & v_y(\mathbf{k})>0 \\
    g_R e^{-\frac{y-L_y}{v_y\tau_0}} + \int_{L_y}^{y} e^{-\frac{y-\xi}{v_y\tau_0}} \frac{1}{v_y\tau_0} \bar{g}(\xi) \,\mathrm{d} \xi, & v_y(\mathbf{k})<0
\end{cases} \label{eq.gsolution}
\end{align}  
\end{widetext}

According to the continuity equation Eq.~\eqref{eq.Iyconser}, we know the electrical current is constant. And by using the linear approximation Eq.~\eqref{eq.gbarlinear}, we derive the expression for \( g(\mathbf{k},y) \) at \( y=\frac{L_y}{2} \) as
\begin{widetext}
\begin{align}\label{eq.gres}
g(\mathbf{k},y=\frac{L_y}{2}) =\frac{1}{L_y+2l_f}[v_y\tau_0(1-e^{-\frac{L_y}{2\mid v_y\mid\tau_0}})+\frac{v_y}{\mid v_y\mid}l_f e^{-\frac{L_y}{2\mid v_y\mid\tau_0}}](g_L-g_R)
\end{align}
\end{widetext}

\begin{figure*}[t]
	\centering
	\includegraphics[width=\linewidth]{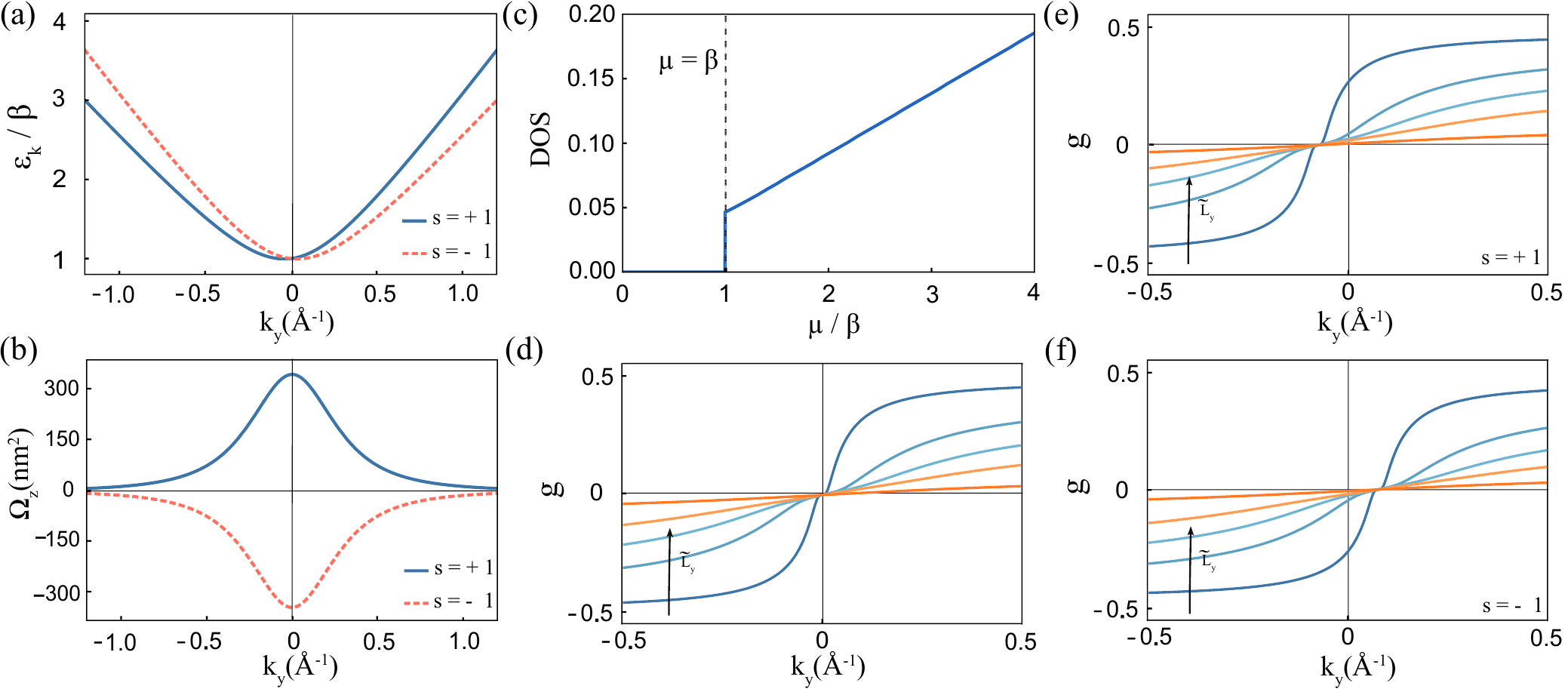}
    \caption{(a) Tilted Dirac cone band structure in the ferroelectric phase of TCI. We consider isotropic Fermi velocity $v_{F}=v_{F x}=v_{F y}=4 \times 10^{5} \mathrm{~m} / \mathrm{s}$  and tilted parameter $\alpha=0.1 v_{F}$ as well as gap $\beta=10 \mathrm{meV}$. (b) Berry curvature of the system. (c) The density of states which $\mu$ is the Fermi energy. (d) The distribution function $g_s(y)$ when the tilted parameter is zero. And $\tilde{L}_y=\frac{L_y}{l_f}$ is defined as a dimensionless quantity. (e) and (f) The distribution function $g_s(y)$ in two energy valleys when $\alpha=0.1 v_{F}$.}\label{fig2}
\end{figure*}

Considering the theory ensures the gauge invariance of the current~\cite{gengUnifiedSemiclassicalApproach2016}, we can adopt the asymmetric gauge with the condition $g_L+g_R=0$. Therefore, we have 
\begin{align}\label{eq.Iyres}
    j_y &=-\frac{2e}{\left(2\pi\right)^2}\int \tilde{v}_y\ f(\mathbf{k},y=\tfrac{L_y}{2})d^{2}k\\ 
&\equiv -\frac{2e}{\left(2\pi\right)^2}\int \tilde{v}_y\ g(\mathbf{k},y=\tfrac{L_y}{2})(-\frac{\partial f_0}{\partial \epsilon_k})d^{2}k
\end{align}
where $\tilde{v}_y = v_y+\frac{e}{\hbar}\frac{\partial V(x)}{\partial x}\Omega_z$ is the group velocity $v_y=\frac{1}{\hbar}\frac{\partial \epsilon_k}{\partial k_y}$ and the anomalous velocity induced by the Berry curvature along the $y$ direction. In our setup, the electric field is oriented along the $y$ direction, thus the anomalous velocity vanishes.
From Eq.~\eqref{eq.Iyres}, we also see that $g(\mathbf{k},y=\tfrac{L_y}{2})(-\frac{\partial f_0}{\partial \epsilon_k})$ and $f(\mathbf{k},y=\tfrac{L_y}{2})$ play the same role in calculating current. The preceding derivation demonstrates that the current is solely determined by the voltage difference between the boundaries of the sample, irrespective of the spatial profile of the electric field $-\frac{\partial V(y)}{\partial y}$.
The detail of the current calculation can be found in our previous work~\cite{gengUnifiedSemiclassicalApproach2016}.

Equipped with the solution of $g(\mathbf{k},y)$ at Eq.~\eqref{eq.gres} and the non-equilibrium distribution function at Eq.~\eqref{eq.fexpand}, we are able to calculate the Hall response of our system. Then, we acquire
\begin{equation}\label{eq.Ixres}
j_x=-\frac{2e}{\left(2\pi\right)^2}\int \tilde{v}_x g(\mathbf{k},y)(-\frac{\partial f_0}{\partial \epsilon_k})d^{2}k~.
\end{equation}
Here, 
\begin{equation}
    \tilde{v}_x=v_x-\frac{e}{\hbar}\frac{\partial V(y)}{\partial y}\Omega_z~,\quad v_x=\frac{1}{\hbar}\frac{\partial \epsilon_k}{\partial k_x}
\end{equation}
is the velocity along the $x$ direction. The normal transverse current induced by the velocity $v_x$ is zero, because the distribution function $f_0$ and $g(\mathbf{k},y)$ are independent of $v_x/\left|v_x\right|$. The normal transervse current is given by
\begin{equation}
\begin{split}
    I_x^0&=\frac{e L_y}{\pi^2}\int v_x\left(f_0+g(\mathbf{k},y)(-\frac{\partial f_0}{\partial \epsilon_k})\right)dk_x dk_y \\
    &=\frac{e L_y}{\pi^2}\int v_x \left(f_0+g(\mathbf{k},y)(-\frac{\partial f_0}{\partial \epsilon_k})\right)\left|1/\frac{\partial \epsilon_k}{\partial k_x}\right|d\epsilon_k dk_y \\
    &=\frac{e L_y}{\pi^2}\int \frac{v_x}{\left|v_x\right|} \left(f_0+g(\mathbf{k},y)(-\frac{\partial f_0}{\partial \epsilon_k})\right)d\epsilon_k dk_y\\
    &=0~,
\end{split}
\end{equation}
for any given $\epsilon_k$, positive and negative moving modes along $k_x$ direction are equal in number. To obtain the Hall current arising from the Berry curvature, we integrate the anomalous velocity contribution over the whole length of the device $I_{x}=\int_{0}^{L_y}  j_{x}d y$, then we can define Hall conductance as $ 
I_x=G_{xy}\left(V_L-V_R\right)^2$ and obtain

\begin{widetext}
\begin{align}
G_{x y}=\frac{ e^{3}}{  \pi h} \frac{1}{L_{y}+2 l_{f}} \int \Omega_{z} \Bigg[ v_{y}\tau_{0} \left(1-e^{-\frac{L_{y}}{2 \left|v_{y}\right|\tau_{0}}}\right)+ \frac{v_{y}}{\left|v_{y}\right|} l_{f} e^{-\frac{L_{y}}{2 \left|v_{y}\right|\tau_{0}}} \Bigg] \left(-\frac{\partial f_{0}}{\partial \epsilon_{k}}\right) d^{2} k~. \label{eq.16} 
\end{align} 
\end{widetext} 

This formula constitutes the central result of our article, providing a universal description of the nonlinear Hall effect at different sizes. By taking the limit of $L_y\gg l_f$ and $L_y\ll l_f$, we can obtain the diffusive and ballistic limits of the nonlinear Hall effect, which correspond to two distinct geometric origins of Hall conductance. More specifically, one originates from Berry curvature dipoles, while the other arises from the integral of the Berry curvature over the Fermi surface.
In the diffusive limit with $L_y\gg l_f$, we obtain
\begin{equation}
G_{xy}^{dif}=\frac{e^{3}}{\pi h} \frac{\tau_{0}}{L_{y}} \int \Omega_{z} v_{y}\left(-\frac{\partial f_{0}}{\partial \epsilon_{k}}\right) d^{2} k,~L_y \gg l_f\label{eq.Gdif} 
\end{equation} 
which precisely matches the established Berry curvature dipole formula\cite{sodemannQuantumNonlinearHall2015b}. 
And in the ballistic limit when $L_y\ll l_f$, we have
\begin{equation}
G_{xy}^{bal}=\frac{e^{3}}{2\pi h} \int \frac{v_{y}}{\left|v_{y}\right|} \Omega_{z}\left(-\frac{\partial f_{0}}{\partial \epsilon_{k}}\right) d^{2} k,~L_y \ll l_f\label{eq.Gbal} 
\end{equation} 
which exhibits a form consistent with the Magnus Hall effect~\cite{papajMagnusHallEffect2019}. This dual consistency conclusively demonstrates our theory  is capable of bridging disparate transport regimes through a single unified formalism.

\section{Size Effect of Nonlinear Hall Effect in the TCI}\label{sec.result} 
In this section, we specifically study the TCI systems to elucidate the size dependence of nonlinear Hall response. The (001) surface of this TCI hosts four massless Dirac fermions, protected by two mirror symmetries\cite{hsiehTopologicalCrystallineInsulators2012, andoTopologicalCrystallineInsulators2015, liuTwoTypesSurface2013, okadaObservationDiracNode2013, serbynSymmetryBreakingLandau2014}. Below the critical temperature $T_c$, a spontaneous structural phase transition only breaks one of the two mirror symmetries in the system. This reduction of symmetry causes two of the four massless Dirac fermions to gain mass. Since the remaining massless Dirac points have vanishing Berry curvature, the Hall current is dominated by the two massive Dirac fermions in the distorted crystal structure. Under time-reversal symmetry \(\hat{T}\) mapping, two massive Dirac fermions acquire Berry curvatures of opposite polarities, a direct manifestation of T-symmetry in momentum space. The low-energy Hamiltonian for the massive Dirac point is given by
\begin{equation}\label{eq.FullHam}
 H(\mathbf{k}) = \begin{pmatrix}
    H_{+1}(\mathbf{k}) & 0 \\
    0 & H_{-1}(\mathbf{k})
\end{pmatrix} 
\end{equation}
where \(H_{s}(\mathbf{k})\) is the Hamiltonian for the two valleys with valley index \(s=\pm1\) and \(\mathbf{k}=(k_x,k_y)\) is the momentum in the two-dimensional. The valley resolved Hamiltonian can be expressed as
\begin{equation}\label{eq.Ham}
    H_{s}(\mathbf{k})=-s k_{y} v_{F y} \sigma_{x}+k_{x} v_{F x} \sigma_{y}+s \alpha k_{y}+\beta \sigma_{z}~,
\end{equation}
where \(v_{F x}\) and \(v_{F y}\) are the Fermi velocities along the \(x\) and \(y\) directions, respectively. The Pauli matrices \(\sigma_{x,y,z}\) act on the spin degree of freedom, and \(\alpha\) is the tilt parameter that breaks the time-reversal symmetry at a single valley. The term \(\beta\) represents the energy gap opened by the ferroelectric distortion, which is a key feature of TCI systems.
The energy dispersion for the Dirac cone is
\begin{align}
    \epsilon_{s}(k)=s\alpha k_{y}+\operatorname{sgn}(\mu) \sqrt{\beta^{2}+k_{x}^{2} v_{F_x}^{2}+k_{y}^{2} v_{F_y}^{2}}~,
\end{align}
where $\mu>0(\mu<0)$ for conduction (valence) band. With the presence of an asymmetric tilt term $\left ( \alpha k_y \right )$, the Dirac cones of the two valleys undergo relative displacement along the $k_y$ direction, resulting in their band structures no longer overlapping. The unique band structure of the system in momentum space is illustrated in Fig.~\ref{fig2}(a). Notably, the valley index $s$ reverses its sign under the time-reversal operation $\hat{T}=s_x \otimes \hat{I}\hat{K}$ (where $s_x$ is the first Pauli matrix acting on the valley space), so a single energy valley violates time-reversal symmetry. However, the entire system satisfies the relationship $\hat{T}H^*(\mathbf{k})\hat{T}^{-1}=H(\mathbf{-k})$. This special symmetry breaking and recovery mechanism ensures that the system maintains time reversal symmetry. The Berry curvature is
\begin{align}
    \Omega_{z,s}=\frac{\operatorname{sgn}(\mu)}{2} \frac{s v_{F x} v_{F y} \beta}{\left(\beta^{2}+k_{x}^{2} v_{F x}^{2}+k_{y}^{2} v_{F y}^{2}\right)^{3 / 2}}~.
\end{align}
\subsection{Symmetry analysis}
Before delving into further details, let us first conduct a symmetry analysis. The transverse current generated by the anomalous velocity arising from the Berry curvature may yield a non-zero value. For time-reversal invariant systems, we can derive the following general relations
\begin{equation}\label{eq.timerev}
    \begin{aligned}
    \epsilon_{k,s}(\mathbf{k})&=\epsilon_{k,-s}(-\mathbf{k})~,\\
    \mathbf{v}_{s}(\mathbf{k})&=-\mathbf{v}_{-s}(-\mathbf{k})~,\\
    \Omega_{z,s}(\mathbf{k})&=-\Omega_{z, -s}(-\mathbf{k})~.
    \end{aligned}
\end{equation}
It can be seen that the Berry curvature $\Omega_{z,s}$ is an odd function of $\mathbf{k}$, so the integral of $\Omega_{z,s} f_0$ is zero. 
From relations of Eq.~\eqref{eq.timerev} and Eq.~\eqref{eq.gres}, we neglect the valley scattering process (which means the relaxation time between the two valleyes is infinity). And we can get that $g_s(\mathbf{k},y)$ is an odd function of $k_y$ and an even function of $k_x$. 
\begin{equation}
 \begin{aligned}
    g_s(-k_x,k_y)&=g_{-s}(-k_x,k_y)~,\\
g_s(k_x,-k_y)&=-g_{-s}(k_x,-k_y)~.
\end{aligned}   
\end{equation}
and in this case the integral of $\sum_{s=\pm 1}\Omega_{z,s} g_s(\mathbf{k},y)$ over the Fermi surface may remain nonvanishing. Note that the single valley of TCI system satisfies the following symmetry
\begin{equation}
 \begin{aligned}
    \epsilon_{s}(k_x,k_y)&\neq\epsilon_s(k_x,-k_y)~,\\
   v_{y,s}(k_x,k_y)&\neq- v_{y,s}(k_x,-k_y)~,\\
    \Omega_{z,s}(k_x,k_y)&=\Omega_{z,s}(-k_x,-k_y)~,\\
     \Omega_{z,s}(k_x,k_y)g_s(k_x,k_y)&\neq -\Omega_{z,s}(k_x,-k_y)g_s(k_x,-k_y)~.
\end{aligned}   
\end{equation}
Given the identical symmetries of the two valleys' energy dispersion relations, group velocities and Berry curvatures, we can conclude that the integral for $\Omega_{z,s} g_s(\mathbf{k},y)$ is non-zero and the contributions of the two energy valleys to conductance are identical. 

\subsection{Size effect and Hall conductance}
In the diffusive limit, the nonlinear Hall conductivity defined by $G_{x y}^{H}=\sigma_{x y}^{H} \frac{1}{L_{y}}$ takes the form of
\begin{align}
    \sigma_{d i f}^{H}=\frac{2 e^{3} \tau_{0}}{(2 \pi)^{2} \hbar} \sum_{s=\pm 1}\int \Omega_{z,s} v_{y,s}\left(-\frac{\partial f_{0}}{\partial \epsilon_{k,s}}\right) d^{2} k\label{eq.btefull2}
\end{align}
where $\left(-\frac{\partial f_{0}}{\partial \epsilon_{k,s}}\right)=\delta\left(\epsilon_{k,s}-\mu\right)$ at $T=0K$. The zero-temperature conductivity calculated by Eq.~\eqref{eq.btefull2} is obtained as
\begin{align}
\sigma_{d i f}^{H}=\frac{3 \alpha \beta e^{3} \tau_{0}}{4 \pi} \frac{\mu^{2}-\beta^{2}}{\mu^{4}}
\end{align}

\begin{figure}[t]
	\centering
	\includegraphics[width=\linewidth]{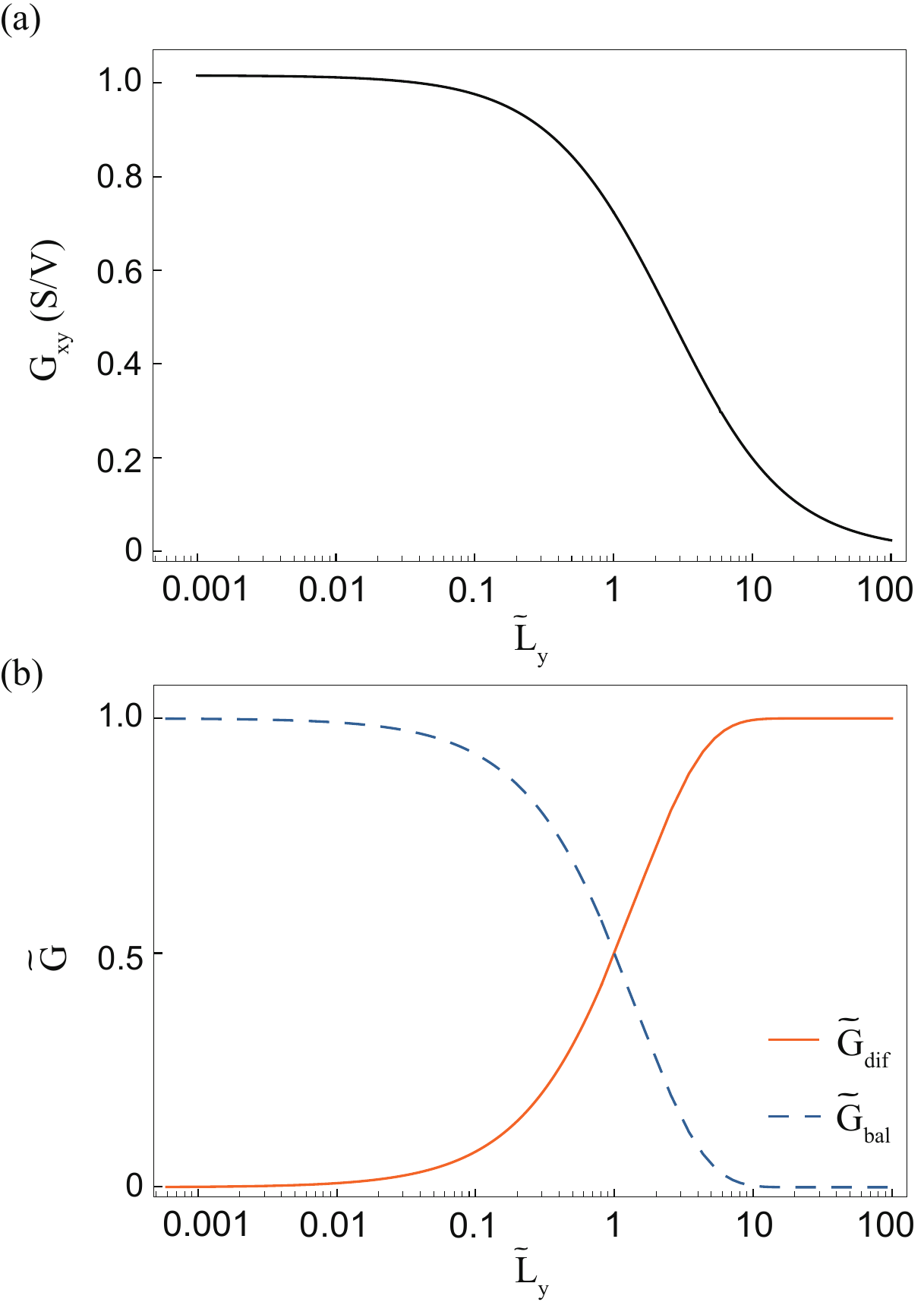}
    \caption{(a) Scaling laws of nonlinear Hall conductance in finite-sized systems. (b) Proportion of contribution of conductance. $\tilde{G}_{dif}$ is the ratio of the diffusive conductance to the total conductance, $\tilde{G}_{bal}$ is the ratio of the ballistic conductance to the total conductance.}\label{fig3}
\end{figure}

The conductance demonstrates distinct scaling behaviors across different size regimes. When the size $L_y$ becomes comparable to the electron mean free path, the conductance exhibits an approximately linear dependence as shown in Fig.~\ref{fig3}(a). This phenomenon arises from the interplay between surface scattering and grain boundary scattering, which reduces the effective mean free path as the system size decreases. In the ballistic transport limit $L_y \ll l_f$, the conductance transcends classical size constraints and manifests constant characteristics. Remarkably, its value is solely determined by the integral of the Berry curvature over the Fermi surface, exhibiting complete independence from geometric dimensions—a finding that aligns perfectly with theoretical predictions of the Magnus Hall effect. The transition behavior between the two regimes can be uniformly described by Eq.~\eqref{eq.16}, where the distribution function essentially reflects the continuous transition from diffusive to ballistic transport as the system size varies. This framework elucidates the transition of charge transport from classical to quantum regimes.

Fig.~\ref{fig3}(b) clearly illustrates the dependence of charge transport mechanisms on system size. As the characteristic size increases, we observe a gradual enhancement in the contribution from diffusive transport becomes more dominant, while ballistic transport diminishes correspondingly. This continuous transition between transport modes reveals how dimensionality fundamentally alters the dominant conduction mechanism. In the macroscopic limit (large system size), where electron scattering processes dominate, the transport becomes entirely diffusive, while the ballistic contribution vanishes completely. Conversely, in the nanoscale limit (extremely small system size), with scattering events becoming negligible, the system exhibits pure ballistic transport without the diffusive contribution. The smooth crossover between these two extremes reveals the intricate interplay between length scales and charge transport physics, where the relative importance of scattering processes and phase coherence effects is modulated by system dimensionality. This behavior can be quantitatively described by our theoretical result in Eq.~\eqref{eq.16}, which seamlessly depicts the whole transition spectrum from ballistic to diffusive regimes by unifying both transport mechanisms. 

\begin{figure}[t]
	\centering
	\includegraphics[width=\linewidth]{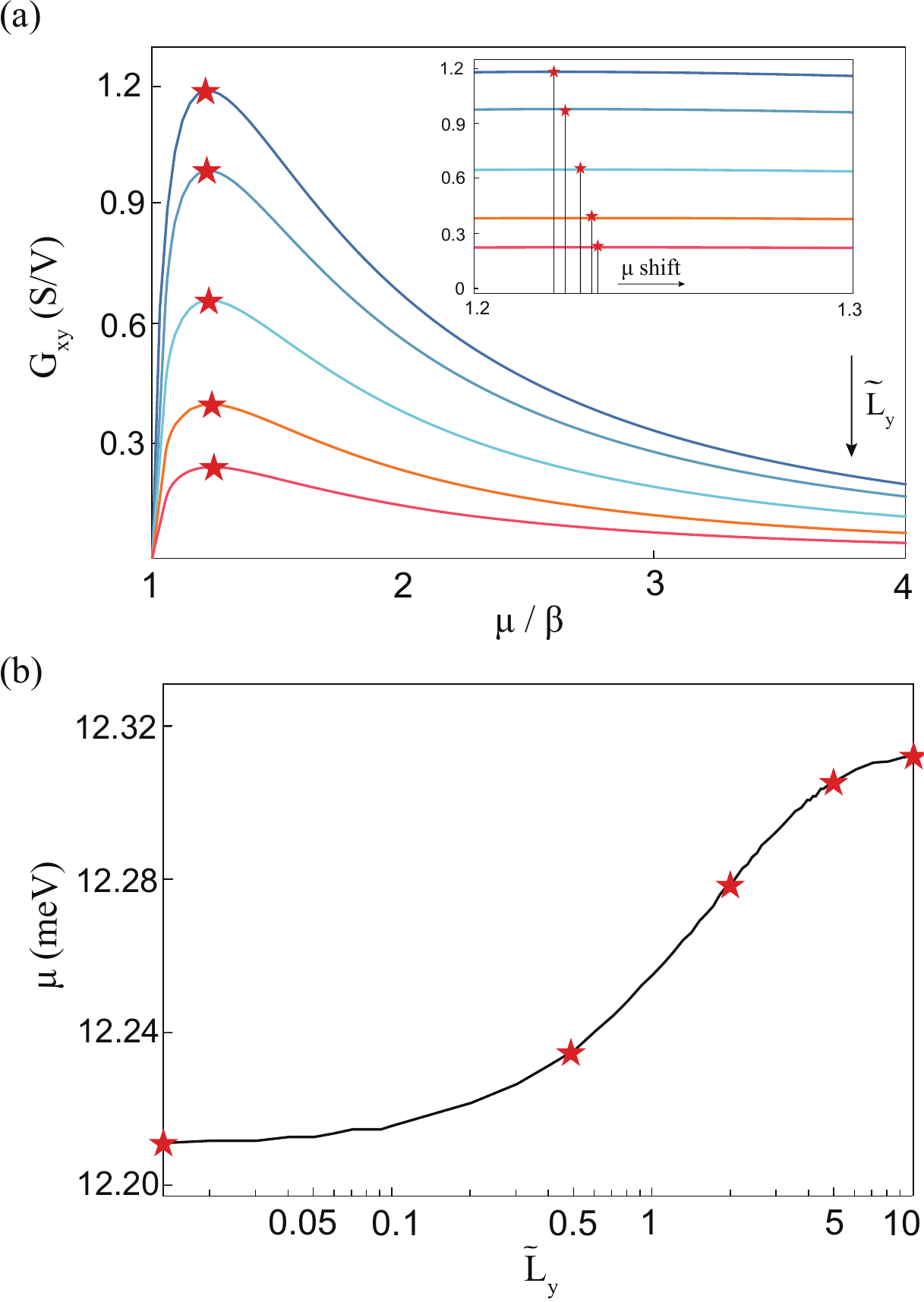}
    \caption{(a) Hall conductance  as a function of the Fermi energy $\mu$ in different sizes (the red point represent where the peak conductance is located). (b) The Fermi energy $ \mu_{max}$ corresponding to the peak Hall conductance varies with size.}\label{fig4}
\end{figure}

Our research indicates that the Hall conductance decays with increasing system size, as shown in Fig.~\ref{fig4}(a). This size dependence stems from synergistic interplay among the density of states near the Fermi level, Berry curvature, and carrier distribution function. Remarkably, the Fermi energy corresponding to the Hall conductance peak has a rightward shift with increasing system size. In Fig.~\ref{fig4}(b), the Fermi energy at the peak of the Hall conductance increases as the system size increases. Through quantitative analysis of Fig.~\ref{fig2}(e) and (f), it is evident that this phenomenon is directly related to the gradual flattening of the slope of the carrier distribution function with increasing system size. This discovery holds significant practical value. By measuring the position of the Hall conductance peak, it is possible to evaluate impurity concentration or carrier relaxation time in the sample, offering a novel experimental approach for material characterization.

\section{\label{sec.conclusion} Conclusion}
This study successfully reveals the size-dependent behavior of charge transport characteristics with size variation in time-reversal symmetric systems by establishing a unified theoretical framework that bridges transport behavior under two distinct limits. We investigated the size effect of nonlinear Hall effect in the TCI and observed that as the system size increases, the Hall conductance undergoes a pronounced and regular decay, while the Fermi energy corresponding to its peak shifts toward higher values. More importantly, this study, for the first time, established a quantitative relationship between the shift of the peak position of Hall conductance and intrinsic material parameters, including impurity concentration and relaxation time. This significant discovery not only enhances our understanding of size-dependent transport mechanisms, but also lays a theoretical foundation for the development of new material characterization techniques. 
We expect that by experimentally measuring the size dependence of Hall conductance, key material parameters such as the tilt parameter and the Berry curvature distribution on the Fermi surface may be estimated, opening new avenues for performance optimization and quality control of nanodevices.

\begin{acknowledgments}
    This work was supported by  the National Natural Science Foundation of China under Grant
    No. 12304068 (H. G.),
    No. 12274235,
    No. 12174182 (D.Y.X.),
    the startup Fund of Nanjing University of Aeronautics and Astronautics Grant No. YAH24076 (H. G.),
    and the State Key Program for Basic Researches of
    China under Grants No. 2021YFA1400403 (D.Y.X.). 
    The computations are partially supported by High Performance Computing Platform of Nanjing University of Aeronautics and Astronautics.
\end{acknowledgments}

\bibliography{ref.bib}

\end{document}